\documentclass[twocolumn,english,aps,prstab]{revtex4-1}

\usepackage{hyperref}
\usepackage{graphicx}

\usepackage{amsmath,amssymb,xfrac}
\usepackage{graphicx}
\usepackage{lipsum}

\newcommand{\mum}{~$\mu$m}

\begin{document}

\title{TeV acceleration in a Matryoshka plasma channel}
\author{J.~P. Farmer and A.~Pukhov}
\affiliation{Institut f\"ur Theoretische Physik I, Universit\"at D\"ussseldorf, 40225
Germany}

\begin{abstract}
Plasma-based accelerators offer accelerating fields orders of magnitude higher than conventional radio-frequency cavities.  However, these accelerating fields are typically coupled with strong focusing fields, which can result in significant radiative energy loss at high energies.  In this work, we present a new accelerator configuration, the Matryoshka channel, a hollow plasma channel stabilized by a hollow plasma filament.  We demonstrate that the absence of on-axis focusing plasma fields greatly reduces radiation losses, allowing acceleration to the TeV frontier.
\end{abstract}

\maketitle

Plasma-based acceleration techniques \cite{lwfa-tajimadawson} continue to garner significant interest due to the high field gradients that they support - orders of magnitude larger than conventional accelerators \cite{lwfa-malka-review,pwfa-hidding-review}.  These high gradients offer the potential to greatly reduce the size of accelerator structures from the kilometre scale to the metre scale.

The premise is simple - a driver, either a charged particle bunch \cite{pwfa-blumenfeld-afterburner} or a laser pulse \cite{lwfa-pukhov-broken,lwfa-faure-dream}, 
passes through the plasma, perturbing the local plasma electron density.  The resulting oscillation in the plasma charge density has a phase velocity equal to the group velocity of the driver.  A witness bunch can be injected such that it suppresses this wakefield, resulting in the witness gaining energy.

For a laser driver, the energy gain is limited by dephasing between the driver and the witness, which may be overcome by staging \cite{lwfa-steinke-staged}.  For a charged-particle driver, the limitation is driver depletion.  This could be overcome by staging, but a more attractive approach is to use a structured driver \cite{pwfa-bane-ramped}.  This allows a larger accelerating field acting on the witness relative to the decelerating field acting on the driver, increasing the energy gain achievable in a single stage.  This transformer ratio is limited to 2 for a symmetric driver \cite{pwfa-ruth-transformer}, and is constrained by the driver length for a tailored beam \cite{pwfa-baturin-accelerating_limit}.

In addition to strong accelerating fields, the plasma density perturbation leads to strong transverse fields.  In many regimes, these fields are beneficial.  In the quasi-linear regime \cite{lwfa-rosenzweig-quasilinear}, there exist separate phases of the wakefield where electrons or positrons may be accelerated and simultaneously focused.  In the blowout regime, betatron oscillations in the focusing field lead to synchrotron emission which may be used as a diagnostic tool \cite{lwfa-corde-betatron_diagnostics}.  Interaction with a laser pulse may be used to enhance this emission \cite{plasma-cipiccia-betatron,lwfa-feng-betatron}.

The radiative losses arising due to these strong focusing fields will ultimately act to reduce the energy gain in a plasma accelerator.  Even in the case where the wakefield generated by the driver is weak, a low-emittance witness beam will drive its own nonlinear wake, again resulting in strong focusing.  An analytical treatment of the blowout regime shows a self-similar scaling, with the radiation-reaction force tending to 2/3 of the accelerating force \cite{lwfa-kostyukov-radiationdamping}.

Acceleration in a hollow plasma channel \cite{plasma-kimura-hollow} would avoid the presence of focusing fields.  This would have the dual benefits of avoiding the modulation instability \cite{pwfa-lotov-selfmodulation,pwfa-kumar-selfmodulation} and minimizing betatron oscillations.  The former allows the use of a long driver, maximizing the transformer ratio, while the latter reduces the radiative energy loss.  Unfortunately, acceleration in a hollow plasma channel is subject to the hosing instability \cite{pwfa-whittum-channel,pwfa-schroeder-channel} in which the driver and witness are attracted to the channel walls, preventing any meaningful acceleration.  Recent work has shown that stable acceleration is possible by using a coaxial plasma channel \cite{pwfa-pukhov-coaxialchannels}, a hollow channel stabilized by an on-axis plasma filament.  This avoids the modulation instability, allowing a high transformer ratio.  However, the strong focusing by the ion filament means that radiation damping at high energies will persist.

In this paper, we present a new configuration -- the Matryoshka channel -- a hollow plasma channel stabilized by a hollow on-axis plasma filament.  A tightly focused drive beam excites a wakefield in the channel and simultaneously scatters electrons from the filament.  The nonlinear focusing provided by the filament leads to transverse phase mixing within the driver. For a sufficiently high transverse momentum spread, this suppress the hosing instability.  A short witness bunch does not require phase mixing to stabilize against the hosing instability, and so a low-emittance witness propagating within the ion filament may be used, almost eliminating radiation damping.

\begin{figure}
\includegraphics{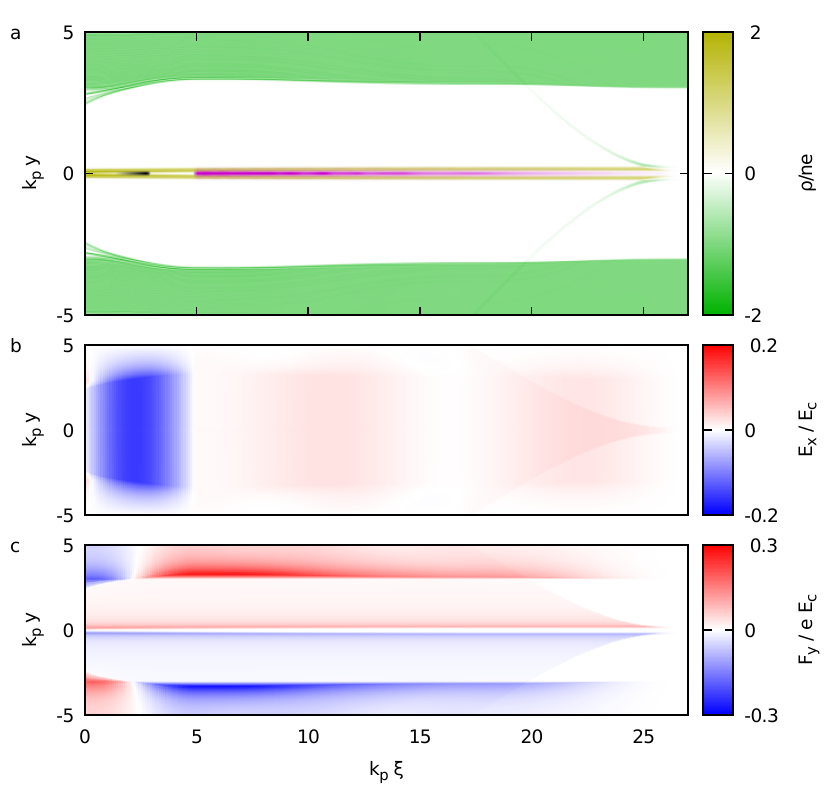}
\caption{\label{fig:geometry}Geometry of the Matryoshka channel.  (a) A ramped drive beam (purple) propagates left-to-right through a hollow plasma filament, which itself lies along the axis of a hollow plasma channel.  The response of the plasma electrons (green) and filament ions (yellow) generates (b) longitudinal and (c) transverse wakefields, which can be used to accelerate a witness bunch inside the filament (shown in black in (a)), while keeping the driver focused.  The plotted density, $\rho$, represents the total charge of the filament plus the electron charge for the bulk plasma.  $E_{c}=mc\omega_{p}/e$ is the critical field.}
\end{figure}

The configuration of the Matryoshka channel, so named because of the Russian doll, is shown in Fig.~\ref{fig:geometry}.  It consists of a hollow plasma channel - a cylindrical void in the bulk plasma - with a narrow, on axis plasma filament, which itself contains a hollow channel.  
A plasma density of $n=1.1\times10^{17}$~cm$^{-3}$ is used for both the bulk plasma and the filament, corresponding to a plasma wavelength $2\pi/k_p=100$\mum , where $k_p=\omega_p/c$.  The plasma channel has a radius of $3/k_p$, and the filament has an outer radius of $0.22/k_p$ and an inner radius of $0.1/k_p$.  In physical units, these correspond to 48, 3.6 and 1.6\mum .  For the driver, a simple linearly ramped current profile is used, with length $22/k_p$ and a peak current of 6.7~kA.  A ramped witness beam of length $1.5/k_p$ and charge 30~pC propagates behind the driver.  Both the driver and witness have an initial energy of 150~GeV.  Helium is used for the filament ions, while ions in the bulk plasma are assumed to be stationary.

In order to demonstrate this scheme, we use the fully three-dimensional quasi-static particle-in-cell code {\texttt qv3d} \cite{pic-pukhov-quasistatic}, developed on the VLPL platform \cite{pic-pukhov-vlpl}.  This greatly reduces the computational overhead of simulations while retaining the relevant physics, allowing acceleration into the TeV regime to be modelled.  
A simulation cell size of $(0.1\times0.04\times0.04)/k_p$ is used, with the plasma particle push divided into two substeps, and a timestep of $100/\omega_p$.  4 particles per cell are used to model the bulk plasma, and 64 for the filament.  For the driver and witness, a near constant macroparticle weight is used within each beam slice, which ensures the resolution is maintained as particles execute betatron oscillations.  For the initial unfocussed drive beam with RMS radius $0.4/k_p$, 512 particles per cell are used at its maximum density on axis, and 4096 for the initial witness of radius $0.07/k_p$.  This corresponds to an average of $\sim120$ and $\sim220$ particles per cell, respectively, 

\begin{figure}
\includegraphics{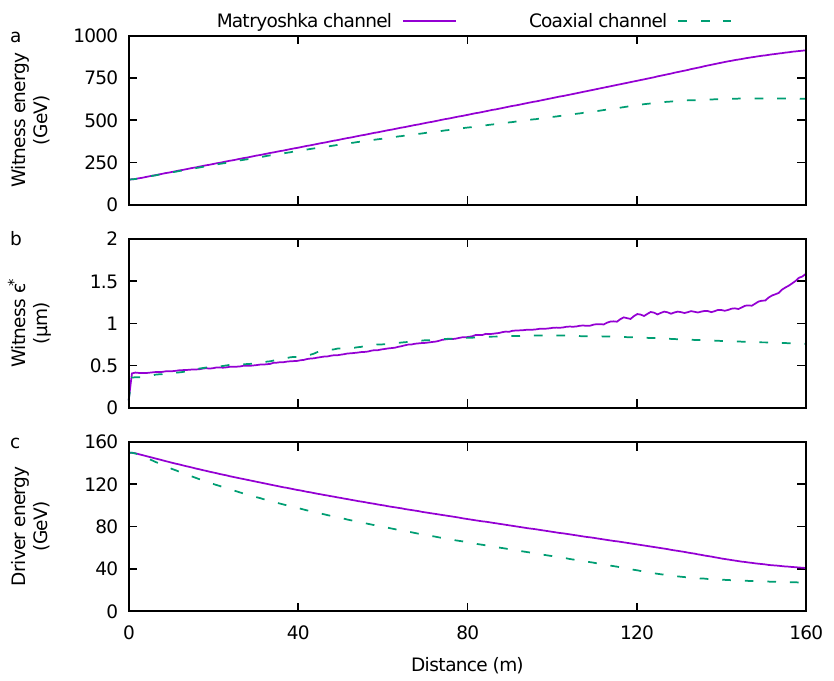}
\caption{\label{fig:accelerate}Evolution of the (a) witness energy and (b) emittance and (c) driver energy, for the Matryoshka (hollow filament, in purple) and coaxial (solid filament, green dashed) channels.  Radiation damping in the coaxial channel prematurely depletes the driver and reduces the witness energy gain, while the Matryoshka channel allows acceleration to the TeV level.  The emittance growth in both cases is likely dominated by numerical noise.}
\end{figure}

Acceleration in the Matryoshka channel is modelled over a distance $1\times10^7/k_p$, corresponding to 160~m.  The evolution of the witness and drive beams are shown in Fig.~\ref{fig:accelerate}.  The witness reaches an energy of 910~GeV, corresponding to a transformer ratio of 5.1.  
The average driver energy falls to 40~GeV, although depletion is nonuniform due to the non-constant decelerating field, as seen in Fig.~\ref{fig:geometry}, resulting in modulation of the driver over this distance.  The normalised witness emittance remains below 1\mum\ over most of the acceleration distance, increasing rapidly towards the end of the simulation as sections of the driver become depleted and dephase.

For comparison, simulations carried out in a coaxial channel \cite{pwfa-pukhov-coaxialchannels}, i.e. with a solid filament, are also shown.  A solid filament of radius $0.2/k_p$ is used to give the same total electron charge as for the hollow filament.  
Other parameters are as for the simulations of the Matryoska channel.  As can be seen, the solid filament leads to slower witness energy gain and faster driver depletion, resulting in a lower final witness energy of 630~GeV.  This corresponds to a transformer ratio of only 3.2, i.e.~a reduction in the energy gain of 40\%, despite the initial wakefield being almost identical to the Matryoshka channel.

To understand this difference, we consider a relativistic electron with Lorentz factor $\gamma$ acted upon by a transverse force $F_\perp$.  The electron will radiate with a power $P_\textrm{rad} = (2r_e/3mc)\gamma^2F_\perp^2$, lowering its total energy.  Here $r_e=(1/4\pi\varepsilon_0)e^2/mc^2$ is the classical electron radius, $-e$ and $m$ are the electron charge and mass, $c$ the vacuum speed of light, and $\varepsilon_0$ is the permittivity of free space.  For betatron oscillations inside a solid ion filament, $\smash{F_\perp=m\omega_\beta^2r=m\omega_p^2r/2\gamma}$, where $r$ is the oscillation amplitude and $\omega_p=\smash{\left(ne^2/\varepsilon_0m\right)^{1/2}}$ is the plasma frequency, with $n$ the quiescent plasma density.

We note that for a focused beam, the normalized emittance reduces to $\epsilon^\ast_y=\gamma\sigma_y\sigma_{y^\prime}$, where $\sigma_{y,y^\prime}$ are the RMS spread of the transverse position and angle of propagation, respectively.  For a relativistic beam undergoing betatron oscillations, this can be written $\epsilon^\ast_y=\gamma\sigma_y\sigma_{v_y}/c=\gamma\omega_\beta\sigma_y^2/c$.  Assuming $\epsilon^\ast_y=\epsilon^\ast_z=\epsilon^\ast$, the average radiated power is therefore
\begin{align}\left<P_\mathrm{rad}\right>=\frac{\sqrt{2}mr_e}{3}\gamma^{3/2}\omega_p^3\epsilon^\ast.
\end{align}
For realistic parameters, this can be a significant fraction of the rate of energy exchange with the wakefield, decreasing the rate at which the witness gains energy, and increasing the rate at which the driver is depleted.

In the Matryoshka channel, the electron beam propagates in a hollow filament of inner radius $r_i$, such that the oscillation amplitude $r$ of an individual electron impinges only a small distance into the filament wall before being reflected.  These half-oscillations inside the wall are approximately sinusoidal in the limit $r-r_i\ll r_i$, with a frequency $\omega_\mathrm{w}=\omega_p/\sqrt{\gamma}$.  Although this frequency is higher than the oscillation frequency $\omega_\beta$ for a solid filament, particles spend only a small time $\tau_\mathrm{w}\sim\pi/\omega_\mathrm{w}$ in the filament wall, compared to the time traversing the cavity, $\tau_c\sim2r_i/((r-r_i)\omega_\mathrm{w})$.  As there are no transverse plasma fields in the cavity, the average radiative loss is therefore greatly reduced.  One additional advantage of this scheme is that $\tau_c$ depends strongly on the oscillation amplitude, introducing significant transverse phase mixing and suppressing the hosing instability.

\begin{figure}
\includegraphics{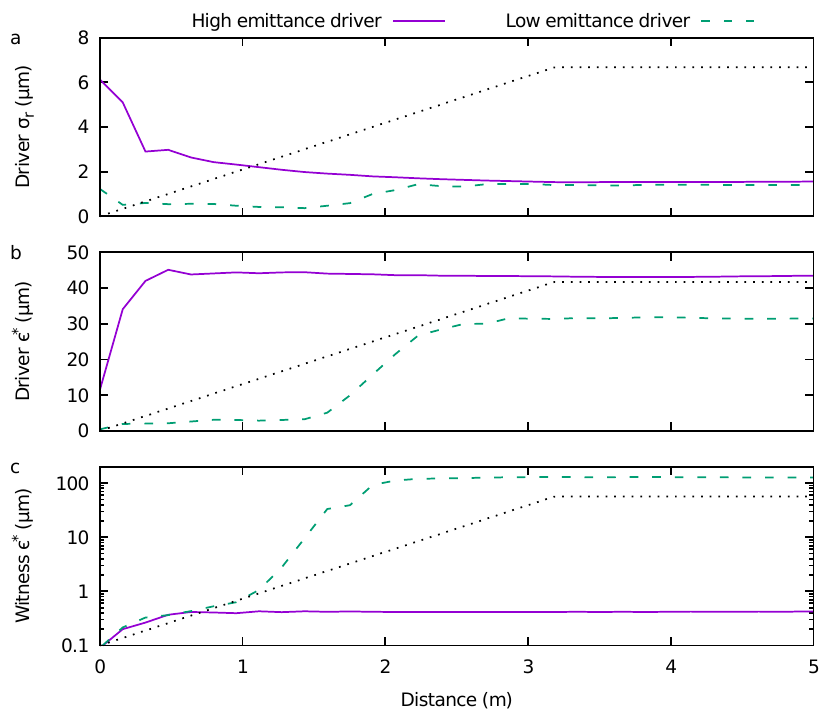}
\caption{\label{fig:thread}Evolution of the driver and witness beams entering the plasma, for both an initially wide (purple) and narrow (green, dashed) driver.  a) driver radius, b) driver emittance, c) witness emittance.  The plasma density is shown (black dotted line) for comparison.  An initially narrow driver will rapidly hose within the filament, greatly increasing the emittance of the witness.}
\end{figure}

To avoid the problem of ``threading the needle'' -- focusing a high emittance drive beam into a narrow hollow filament -- we make use of a plasma density ramp.  An initially wide drive beam with RMS radius $0.4/k_p=6.4$\mum\ is used, with emittance matched to an external quadrupole focusing field gradient of $0.01$~$k_p$T, equivalent to 0.16~T/mm.  Electrons oscillating in the external field become trapped near the axis as the plasma density increases.  A ramp length of $2\times10^4/k_p=3.2$~m is used to allow efficient trapping of the beam.  Figures~\ref{fig:thread}a,b show the evolution of the drive beam as it becomes trapped, with the radius decreasing as the emittance grows.  The resulting phase mixing suppresses the hosing instability, which keeps the witness emittance low, as seen in Fig.~\ref{fig:thread}.  For comparison, an initially narrow driver of radius $0.08/k_p$ matched to the same external field will rapidly hose inside the filament, resulting in an orders-of-magnitude greater witness emittance.  Using an external magnetic field has the additional benefit of preventing the free expansion of the leading edge of the driver, which can modify the wakefield over long distances \cite{pwfa-pukhov-coaxialchannels}.

The creation of the Matryoshka channel could be achieved via ionization by the transverse field of a relativistic charged particle beam, which naturally produces a hollow filament \cite{plasma-tarkeshian-ionizationprofiling}.  This would avoid the need for laser ionization, which would introduce additional complications with alignment.  For a symmetric beam, the on-axis field is zero, and so ionization naturally leads to a hollow filament.  In atomic units, the critical field of ionizition is $E_\mathrm{i}=1/{16{n^\ast}^4}$, with \smash{$n^\ast=Z/\sqrt{2I_p}$} the effective quantum number, where $Z$ is the atomic residue and $I_p$ the ionization potential \cite{plasma-bekov-Rb_ionization}.  Considering the transverse field generated by a Gaussian beam of RMS radius $\sigma_r$ and current $I_b$, we derive an engineering formula for the maximum ionization potential for which the critical ionization field for single ionization can be reached:
\begin{equation}
  I_p\,[\textrm{eV}]\,= 12.6\sqrt{\frac{I_b\,[\mathrm{kA}]}{\sigma_r\,[\mu\mathrm{m}]}}.
\end{equation}
For the parameters used in this work, this corresponds to a required beam current $\sim9.7$~kA for the creation of the hollow filament, and $\sim240$~kA for the outer plasma channel.  The entire process of channel creation and acceleration could be realised in a single beamline.

The high emittance seen in simulations is dominated by numerical noise -- relatively few particles per cell are used, and the particle beams are initialized with a random Gaussian velocity distribution.  Simulations are, however, demanding, and using more particles per cell quickly becomes computationally prohibitive.  In order to reduce the computational overhead, allowing a higher resolution to be used, a fully 3D cylindrical-geometry model for {\texttt qv3d} is under active development, and will be published elsewhere.  Furthermore, a number of physical techniques may be used to reduce the emittance.  Phase mixing reduces the growth rate of the hosing instability, and so a higher driver emittance could be used to further improve the quality of the witness beam.  Adding an energy chirp to the driver could also reduce the witness emittance through the BNS stabilization mechanism \cite{pwfa-BNS}.  If the emittance could be reduced to the order of 0.1\mum, it would also be possible to accelerate positrons in the channel, with the bunch constrained by an external quadrupole field.

Further optimization of the system is certainly possible.  A tailored drive beam can be used to give a near-constant accelerating field acting on the driver, maximizing the transformer ratio \cite{pwfa-bane-ramped}.  A full treatment requires that the nonlinear response of the channel be taken into account \cite{pwfa-farmer-channel_loading}. Use of a tailored witness current profile allows the accelerating field to be locally flattened through beamloading, resulting in a monoenergetic bunch \cite{lwfa-Couperus-loaded}.

The main constraints of the Matryoshka channel are on the driver.  The driver current should not be so high as to cause additional ionization, which would act to cause additional beamloading and prematurely deplete the pump.  Further, the integrated driver charge should not be so high as to cause the hollow ion filament to collapse before the witness, which would introduce radiative losses as in the solid filament case.  The driver emittance also needs to be sufficiently large to prevent the hosing instability from developing, which in turn would lead to a high witness emittance.

In conclusion, we have demonstrated a new acceleration scheme, the Matryoshka channel, consisting of a hollow plasma channel stabilized by a hollow plasma filament.  The use of a hollow plasma channel allows a long driver to be used, increasing the energy gain achievable in a single stage.  Stabilization by a hollow filament greatly reduces losses to radiation damping, which act to prematurely deplete the driver and lower the rate of energy gain by the witness.  Three dimensional PIC simulations have been carried out to demonstrate the scheme, showing acceleration of a witness bunch from 150 to 910~GeV over 160~m by a 150~GeV driver, representing a transformer ratio of 5.1.  Simulations carried out with the equivalent geometry with a solid filament achieved a transformer ratio of only 3.2.  The Matryoshka channel therefore represents a promising pathway to accelerate electrons to the TeV level using currently available technology.

This work has been supported by the Deutsche Forschungsgemeinschaft and by BMBF.


\begin{thebibliography}{28}%
\makeatletter
\providecommand \@ifxundefined [1]{%
 \@ifx{#1\undefined}
}%
\providecommand \@ifnum [1]{%
 \ifnum #1\expandafter \@firstoftwo
 \else \expandafter \@secondoftwo
 \fi
}%
\providecommand \@ifx [1]{%
 \ifx #1\expandafter \@firstoftwo
 \else \expandafter \@secondoftwo
 \fi
}%
\providecommand \natexlab [1]{#1}%
\providecommand \enquote  [1]{``#1''}%
\providecommand \bibnamefont  [1]{#1}%
\providecommand \bibfnamefont [1]{#1}%
\providecommand \citenamefont [1]{#1}%
\providecommand \href@noop [0]{\@secondoftwo}%
\providecommand \href [0]{\begingroup \@sanitize@url \@href}%
\providecommand \@href[1]{\@@startlink{#1}\@@href}%
\providecommand \@@href[1]{\endgroup#1\@@endlink}%
\providecommand \@sanitize@url [0]{\catcode `\\12\catcode `\$12\catcode
  `\&12\catcode `\#12\catcode `\^12\catcode `\_12\catcode `\%12\relax}%
\providecommand \@@startlink[1]{}%
\providecommand \@@endlink[0]{}%
\providecommand \url  [0]{\begingroup\@sanitize@url \@url }%
\providecommand \@url [1]{\endgroup\@href {#1}{\urlprefix }}%
\providecommand \urlprefix  [0]{URL }%
\providecommand \Eprint [0]{\href }%
\providecommand \doibase [0]{http://dx.doi.org/}%
\providecommand \selectlanguage [0]{\@gobble}%
\providecommand \bibinfo  [0]{\@secondoftwo}%
\providecommand \bibfield  [0]{\@secondoftwo}%
\providecommand \translation [1]{[#1]}%
\providecommand \BibitemOpen [0]{}%
\providecommand \bibitemStop [0]{}%
\providecommand \bibitemNoStop [0]{.\EOS\space}%
\providecommand \EOS [0]{\spacefactor3000\relax}%
\providecommand \BibitemShut  [1]{\csname bibitem#1\endcsname}%
\let\auto@bib@innerbib\@empty
\bibitem [{\citenamefont {Tajima}\ and\ \citenamefont
  {Dawson}(1979)}]{lwfa-tajimadawson}%
  \BibitemOpen
  \bibfield  {author} {\bibinfo {author} {\bibfnamefont {T.}~\bibnamefont
  {Tajima}}\ and\ \bibinfo {author} {\bibfnamefont {J.~M.}\ \bibnamefont
  {Dawson}},\ }\href {\doibase 10.1103/PhysRevLett.43.267} {\bibfield
  {journal} {\bibinfo  {journal} {Phys. Rev. Lett.}\ }\textbf {\bibinfo
  {volume} {43}},\ \bibinfo {pages} {267} (\bibinfo {year} {1979})}\BibitemShut
  {NoStop}%
\bibitem [{\citenamefont {Malka}(2012)}]{lwfa-malka-review}%
  \BibitemOpen
  \bibfield  {author} {\bibinfo {author} {\bibfnamefont {V.}~\bibnamefont
  {Malka}},\ }\href {\doibase http://dx.doi.org/10.1063/1.3695389} {\bibfield
  {journal} {\bibinfo  {journal} {Physics of Plasmas}\ }\textbf {\bibinfo
  {volume} {19}},\ \bibinfo {eid} {055501} (\bibinfo {year}
  {2012})}\BibitemShut {NoStop}%
\bibitem [{\citenamefont {Hidding}\ \emph {et~al.}(2019)\citenamefont
  {Hidding}, \citenamefont {Foster}, \citenamefont {Hogan}, \citenamefont
  {Muggli},\ and\ \citenamefont {Rosenzweig}}]{pwfa-hidding-review}%
  \BibitemOpen
  \bibfield  {author} {\bibinfo {author} {\bibfnamefont {B.}~\bibnamefont
  {Hidding}}, \bibinfo {author} {\bibfnamefont {B.}~\bibnamefont {Foster}},
  \bibinfo {author} {\bibfnamefont {M.~J.}\ \bibnamefont {Hogan}}, \bibinfo
  {author} {\bibfnamefont {P.}~\bibnamefont {Muggli}}, \ and\ \bibinfo {author}
  {\bibfnamefont {J.~B.}\ \bibnamefont {Rosenzweig}},\ }\href {\doibase
  10.1098/rsta.2019.0215} {\bibfield  {journal} {\bibinfo  {journal}
  {Philosophical Transactions of the Royal Society A: Mathematical, Physical
  and Engineering Sciences}\ }\textbf {\bibinfo {volume} {377}},\ \bibinfo
  {pages} {20190215} (\bibinfo {year} {2019})}
  \BibitemShut {NoStop}%
\bibitem [{\citenamefont {Blumenfeld}\ \emph {et~al.}(2007)\citenamefont
  {Blumenfeld}, \citenamefont {Clayton}, \citenamefont {Decker}, \citenamefont
  {Hogan}, \citenamefont {Huang}, \citenamefont {Ischebeck}, \citenamefont
  {Iverson}, \citenamefont {Joshi}, \citenamefont {Katsouleas}, \citenamefont
  {Kirby}, \citenamefont {Lu}, \citenamefont {Marsh}, \citenamefont {Mori},
  \citenamefont {Muggli}, \citenamefont {Oz}, \citenamefont {Siemann},
  \citenamefont {Walz},\ and\ \citenamefont
  {Zhou}}]{pwfa-blumenfeld-afterburner}%
  \BibitemOpen
  \bibfield  {author} {\bibinfo {author} {\bibfnamefont {I.}~\bibnamefont
  {Blumenfeld}}, \bibinfo {author} {\bibfnamefont {C.~E.}\ \bibnamefont
  {Clayton}}, \bibinfo {author} {\bibfnamefont {F.-J.}\ \bibnamefont {Decker}},
  \bibinfo {author} {\bibfnamefont {M.~J.}\ \bibnamefont {Hogan}}, \bibinfo
  {author} {\bibfnamefont {C.}~\bibnamefont {Huang}}, \bibinfo {author}
  {\bibfnamefont {R.}~\bibnamefont {Ischebeck}}, \bibinfo {author}
  {\bibfnamefont {R.}~\bibnamefont {Iverson}}, \bibinfo {author} {\bibfnamefont
  {C.}~\bibnamefont {Joshi}}, \bibinfo {author} {\bibfnamefont
  {T.}~\bibnamefont {Katsouleas}}, \bibinfo {author} {\bibfnamefont
  {N.}~\bibnamefont {Kirby}}, \bibinfo {author} {\bibfnamefont
  {W.}~\bibnamefont {Lu}}, \bibinfo {author} {\bibfnamefont {K.~A.}\
  \bibnamefont {Marsh}}, \bibinfo {author} {\bibfnamefont {W.~B.}\ \bibnamefont
  {Mori}}, \bibinfo {author} {\bibfnamefont {P.}~\bibnamefont {Muggli}},
  \bibinfo {author} {\bibfnamefont {E.}~\bibnamefont {Oz}}, \bibinfo {author}
  {\bibfnamefont {R.~H.}\ \bibnamefont {Siemann}}, \bibinfo {author}
  {\bibfnamefont {D.}~\bibnamefont {Walz}}, \ and\ \bibinfo {author}
  {\bibfnamefont {M.}~\bibnamefont {Zhou}},\ }\href
  {http://dx.doi.org/10.1038/nature05538} {\bibfield  {journal} {\bibinfo
  {journal} {Nature}\ }\textbf {\bibinfo {volume} {445}},\ \bibinfo {pages}
  {741} (\bibinfo {year} {2007})}\BibitemShut {NoStop}%
\bibitem [{\citenamefont {Pukhov}\ and\ \citenamefont {Meyer-ter
  Vehn}(2002)}]{lwfa-pukhov-broken}%
  \BibitemOpen
  \bibfield  {author} {\bibinfo {author} {\bibfnamefont {A.}~\bibnamefont
  {Pukhov}}\ and\ \bibinfo {author} {\bibfnamefont {J.}~\bibnamefont {Meyer-ter
  Vehn}},\ }\href {\doibase 10.1007/s003400200795} {\bibfield  {journal}
  {\bibinfo  {journal} {Applied Physics B}\ }\textbf {\bibinfo {volume} {74}},\
  \bibinfo {pages} {355} (\bibinfo {year} {2002})}\BibitemShut {NoStop}%
\bibitem [{\citenamefont {Faure}\ \emph {et~al.}(2004)\citenamefont {Faure},
  \citenamefont {Glinec}, \citenamefont {Pukhov}, \citenamefont {Kiselev},
  \citenamefont {Gordienko}, \citenamefont {Lefebvre}, \citenamefont
  {Rousseau}, \citenamefont {Burgy},\ and\ \citenamefont
  {Malka}}]{lwfa-faure-dream}%
  \BibitemOpen
  \bibfield  {author} {\bibinfo {author} {\bibfnamefont {J.}~\bibnamefont
  {Faure}}, \bibinfo {author} {\bibfnamefont {Y.}~\bibnamefont {Glinec}},
  \bibinfo {author} {\bibfnamefont {A.}~\bibnamefont {Pukhov}}, \bibinfo
  {author} {\bibfnamefont {S.}~\bibnamefont {Kiselev}}, \bibinfo {author}
  {\bibfnamefont {S.}~\bibnamefont {Gordienko}}, \bibinfo {author}
  {\bibfnamefont {E.}~\bibnamefont {Lefebvre}}, \bibinfo {author}
  {\bibfnamefont {J.-P.}\ \bibnamefont {Rousseau}}, \bibinfo {author}
  {\bibfnamefont {F.}~\bibnamefont {Burgy}}, \ and\ \bibinfo {author}
  {\bibfnamefont {V.}~\bibnamefont {Malka}},\ }\href {\doibase
  10.1038/nature02963} {\bibfield  {journal} {\bibinfo  {journal} {Nature}\
  }\textbf {\bibinfo {volume} {431}},\ \bibinfo {pages} {541} (\bibinfo {year}
  {2004})}\BibitemShut {NoStop}%
\bibitem [{\citenamefont {Steinke}\ \emph {et~al.}(2016)\citenamefont
  {Steinke}, \citenamefont {van Tilborg}, \citenamefont {Benedetti},
  \citenamefont {Geddes}, \citenamefont {Schroeder}, \citenamefont {Daniels},
  \citenamefont {Swanson}, \citenamefont {Gonsalves}, \citenamefont {Nakamura},
  \citenamefont {Matlis}, \citenamefont {Shaw}, \citenamefont {Esarey},\ and\
  \citenamefont {Leemans}}]{lwfa-steinke-staged}%
  \BibitemOpen
  \bibfield  {author} {\bibinfo {author} {\bibfnamefont {S.}~\bibnamefont
  {Steinke}}, \bibinfo {author} {\bibfnamefont {J.}~\bibnamefont {van
  Tilborg}}, \bibinfo {author} {\bibfnamefont {C.}~\bibnamefont {Benedetti}},
  \bibinfo {author} {\bibfnamefont {C.~G.~R.}\ \bibnamefont {Geddes}}, \bibinfo
  {author} {\bibfnamefont {C.~B.}\ \bibnamefont {Schroeder}}, \bibinfo {author}
  {\bibfnamefont {J.}~\bibnamefont {Daniels}}, \bibinfo {author} {\bibfnamefont
  {K.~K.}\ \bibnamefont {Swanson}}, \bibinfo {author} {\bibfnamefont {A.~J.}\
  \bibnamefont {Gonsalves}}, \bibinfo {author} {\bibfnamefont {K.}~\bibnamefont
  {Nakamura}}, \bibinfo {author} {\bibfnamefont {N.~H.}\ \bibnamefont
  {Matlis}}, \bibinfo {author} {\bibfnamefont {B.~H.}\ \bibnamefont {Shaw}},
  \bibinfo {author} {\bibfnamefont {E.}~\bibnamefont {Esarey}}, \ and\ \bibinfo
  {author} {\bibfnamefont {W.~P.}\ \bibnamefont {Leemans}},\ }\href
  {http://dx.doi.org/10.1038/nature16525} {\bibfield  {journal} {\bibinfo
  {journal} {Nature}\ }\textbf {\bibinfo {volume} {530}},\ \bibinfo {pages}
  {190} (\bibinfo {year} {2016})}\BibitemShut {NoStop}%
\bibitem [{\citenamefont {Bane}\ \emph {et~al.}(1985)\citenamefont {Bane},
  \citenamefont {Chen},\ and\ \citenamefont {Wilson}}]{pwfa-bane-ramped}%
  \BibitemOpen
  \bibfield  {author} {\bibinfo {author} {\bibfnamefont {K.~L.~F.}\
  \bibnamefont {Bane}}, \bibinfo {author} {\bibfnamefont {P.}~\bibnamefont
  {Chen}}, \ and\ \bibinfo {author} {\bibfnamefont {P.~B.}\ \bibnamefont
  {Wilson}},\ }\href {\doibase 10.1109/TNS.1985.4334416} {\bibfield  {journal}
  {\bibinfo  {journal} {IEEE Transactions on Nuclear Science}\ }\textbf
  {\bibinfo {volume} {32}},\ \bibinfo {pages} {3524} (\bibinfo {year}
  {1985})},\ \bibinfo {note} {also published in
  \href{https://www.slac.stanford.edu/pubs/slacpubs/3500/slac-pub-3662.pdf}{SLAC-PUB
  3662 (1985) }.}\BibitemShut {Stop}%
\bibitem [{\citenamefont {Ruth}\ \emph {et~al.}(1985)\citenamefont {Ruth},
  \citenamefont {Chao}, \citenamefont {Morton},\ and\ \citenamefont
  {Wilson}}]{pwfa-ruth-transformer}%
  \BibitemOpen
  \bibfield  {author} {\bibinfo {author} {\bibfnamefont {R.~D.}\ \bibnamefont
  {Ruth}}, \bibinfo {author} {\bibfnamefont {A.~W.}\ \bibnamefont {Chao}},
  \bibinfo {author} {\bibfnamefont {P.~L.}\ \bibnamefont {Morton}}, \ and\
  \bibinfo {author} {\bibfnamefont {P.~B.}\ \bibnamefont {Wilson}},\
  }\href@noop {} {\bibfield  {journal} {\bibinfo  {journal} {Particle
  Accelerators}\ }\textbf {\bibinfo {volume} {17}},\ \bibinfo {pages} {171}
  (\bibinfo {year} {1985})}\BibitemShut {NoStop}%
\bibitem [{\citenamefont {Baturin}\ and\ \citenamefont
  {Zholents}(2017)}]{pwfa-baturin-accelerating_limit}%
  \BibitemOpen
  \bibfield  {author} {\bibinfo {author} {\bibfnamefont {S.~S.}\ \bibnamefont
  {Baturin}}\ and\ \bibinfo {author} {\bibfnamefont {A.}~\bibnamefont
  {Zholents}},\ }\href {\doibase 10.1103/PhysRevAccelBeams.20.061302}
  {\bibfield  {journal} {\bibinfo  {journal} {Phys. Rev. Accel. Beams}\
  }\textbf {\bibinfo {volume} {20}},\ \bibinfo {pages} {061302} (\bibinfo
  {year} {2017})}\BibitemShut {NoStop}%
\bibitem [{\citenamefont {Rosenzweig}\ \emph {et~al.}(2010)\citenamefont
  {Rosenzweig}, \citenamefont {Andonian}, \citenamefont {Ferrario},
  \citenamefont {Muggli}, \citenamefont {Williams}, \citenamefont {Yakimenko},\
  and\ \citenamefont {Xuan}}]{lwfa-rosenzweig-quasilinear}%
  \BibitemOpen
  \bibfield  {author} {\bibinfo {author} {\bibfnamefont {J.~B.}\ \bibnamefont
  {Rosenzweig}}, \bibinfo {author} {\bibfnamefont {G.}~\bibnamefont
  {Andonian}}, \bibinfo {author} {\bibfnamefont {M.}~\bibnamefont {Ferrario}},
  \bibinfo {author} {\bibfnamefont {P.}~\bibnamefont {Muggli}}, \bibinfo
  {author} {\bibfnamefont {O.}~\bibnamefont {Williams}}, \bibinfo {author}
  {\bibfnamefont {V.}~\bibnamefont {Yakimenko}}, \ and\ \bibinfo {author}
  {\bibfnamefont {K.}~\bibnamefont {Xuan}},\ }\href {\doibase
  10.1063/1.3520373} {\bibfield  {journal} {\bibinfo  {journal} {AIP Conference
  Proceedings}\ }\textbf {\bibinfo {volume} {1299}},\ \bibinfo {pages} {500}
  (\bibinfo {year} {2010})},\ \Eprint
  {http://arxiv.org/abs/https://aip.scitation.org/doi/pdf/10.1063/1.3520373}
  {https://aip.scitation.org/doi/pdf/10.1063/1.3520373} \BibitemShut {NoStop}%
\bibitem [{\citenamefont {Corde}\ \emph {et~al.}(2012)\citenamefont {Corde},
  \citenamefont {Thaury}, \citenamefont {Phuoc}, \citenamefont {Lifschitz},
  \citenamefont {Lambert}, \citenamefont {Lundh}, \citenamefont {Brijesh},
  \citenamefont {Arantchuk}, \citenamefont {Sebban}, \citenamefont {Rousse},
  \citenamefont {Faure},\ and\ \citenamefont
  {Malka}}]{lwfa-corde-betatron_diagnostics}%
  \BibitemOpen
  \bibfield  {author} {\bibinfo {author} {\bibfnamefont {S.}~\bibnamefont
  {Corde}}, \bibinfo {author} {\bibfnamefont {C.}~\bibnamefont {Thaury}},
  \bibinfo {author} {\bibfnamefont {K.~T.}\ \bibnamefont {Phuoc}}, \bibinfo
  {author} {\bibfnamefont {A.}~\bibnamefont {Lifschitz}}, \bibinfo {author}
  {\bibfnamefont {G.}~\bibnamefont {Lambert}}, \bibinfo {author} {\bibfnamefont
  {O.}~\bibnamefont {Lundh}}, \bibinfo {author} {\bibfnamefont
  {P.}~\bibnamefont {Brijesh}}, \bibinfo {author} {\bibfnamefont
  {L.}~\bibnamefont {Arantchuk}}, \bibinfo {author} {\bibfnamefont
  {S.}~\bibnamefont {Sebban}}, \bibinfo {author} {\bibfnamefont
  {A.}~\bibnamefont {Rousse}}, \bibinfo {author} {\bibfnamefont
  {J.}~\bibnamefont {Faure}}, \ and\ \bibinfo {author} {\bibfnamefont
  {V.}~\bibnamefont {Malka}},\ }\href {\doibase 10.1088/0741-3335/54/12/124023}
  {\bibfield  {journal} {\bibinfo  {journal} {Plasma Physics and Controlled
  Fusion}\ }\textbf {\bibinfo {volume} {54}},\ \bibinfo {pages} {124023}
  (\bibinfo {year} {2012})}\BibitemShut {NoStop}%
\bibitem [{\citenamefont {Cipiccia}\ \emph {et~al.}(2011)\citenamefont
  {Cipiccia}, \citenamefont {Islam}, \citenamefont {Ersfeld}, \citenamefont
  {Shanks}, \citenamefont {Brunetti}, \citenamefont {Vieux}, \citenamefont
  {Yang}, \citenamefont {Issac}, \citenamefont {Wiggins}, \citenamefont
  {Welsh}, \citenamefont {Anania}, \citenamefont {Maneuski}, \citenamefont
  {Montgomery}, \citenamefont {Smith}, \citenamefont {Hoek}, \citenamefont
  {Hamilton}, \citenamefont {Lemos}, \citenamefont {Symes}, \citenamefont
  {Rajeev}, \citenamefont {Shea}, \citenamefont {Dias},\ and\ \citenamefont
  {Jaroszynski}}]{plasma-cipiccia-betatron}%
  \BibitemOpen
  \bibfield  {author} {\bibinfo {author} {\bibfnamefont {S.}~\bibnamefont
  {Cipiccia}}, \bibinfo {author} {\bibfnamefont {M.~R.}\ \bibnamefont {Islam}},
  \bibinfo {author} {\bibfnamefont {B.}~\bibnamefont {Ersfeld}}, \bibinfo
  {author} {\bibfnamefont {R.~P.}\ \bibnamefont {Shanks}}, \bibinfo {author}
  {\bibfnamefont {E.}~\bibnamefont {Brunetti}}, \bibinfo {author}
  {\bibfnamefont {G.}~\bibnamefont {Vieux}}, \bibinfo {author} {\bibfnamefont
  {X.}~\bibnamefont {Yang}}, \bibinfo {author} {\bibfnamefont {R.~C.}\
  \bibnamefont {Issac}}, \bibinfo {author} {\bibfnamefont {S.~M.}\ \bibnamefont
  {Wiggins}}, \bibinfo {author} {\bibfnamefont {G.~H.}\ \bibnamefont {Welsh}},
  \bibinfo {author} {\bibfnamefont {M.-P.}\ \bibnamefont {Anania}}, \bibinfo
  {author} {\bibfnamefont {D.}~\bibnamefont {Maneuski}}, \bibinfo {author}
  {\bibfnamefont {R.}~\bibnamefont {Montgomery}}, \bibinfo {author}
  {\bibfnamefont {G.}~\bibnamefont {Smith}}, \bibinfo {author} {\bibfnamefont
  {M.}~\bibnamefont {Hoek}}, \bibinfo {author} {\bibfnamefont {D.~J.}\
  \bibnamefont {Hamilton}}, \bibinfo {author} {\bibfnamefont {N.~R.~C.}\
  \bibnamefont {Lemos}}, \bibinfo {author} {\bibfnamefont {D.}~\bibnamefont
  {Symes}}, \bibinfo {author} {\bibfnamefont {P.~P.}\ \bibnamefont {Rajeev}},
  \bibinfo {author} {\bibfnamefont {V.~O.}\ \bibnamefont {Shea}}, \bibinfo
  {author} {\bibfnamefont {J.~M.}\ \bibnamefont {Dias}}, \ and\ \bibinfo
  {author} {\bibfnamefont {D.~A.}\ \bibnamefont {Jaroszynski}},\ }\href
  {http://dx.doi.org/10.1038/nphys2090} {\bibfield  {journal} {\bibinfo
  {journal} {Nat Phys}\ }\textbf {\bibinfo {volume} {7}},\ \bibinfo {pages}
  {867} (\bibinfo {year} {2011})}\BibitemShut {NoStop}%
\bibitem [{\citenamefont {Feng}\ \emph {et~al.}(2019)\citenamefont {Feng},
  \citenamefont {Li}, \citenamefont {Wang}, \citenamefont {Li}, \citenamefont
  {Li}, \citenamefont {Yan}, \citenamefont {Wang},\ and\ \citenamefont
  {Chen}}]{lwfa-feng-betatron}%
  \BibitemOpen
  \bibfield  {author} {\bibinfo {author} {\bibfnamefont {J.}~\bibnamefont
  {Feng}}, \bibinfo {author} {\bibfnamefont {Y.}~\bibnamefont {Li}}, \bibinfo
  {author} {\bibfnamefont {J.}~\bibnamefont {Wang}}, \bibinfo {author}
  {\bibfnamefont {D.}~\bibnamefont {Li}}, \bibinfo {author} {\bibfnamefont
  {F.}~\bibnamefont {Li}}, \bibinfo {author} {\bibfnamefont {W.}~\bibnamefont
  {Yan}}, \bibinfo {author} {\bibfnamefont {W.}~\bibnamefont {Wang}}, \ and\
  \bibinfo {author} {\bibfnamefont {L.}~\bibnamefont {Chen}},\ }\href {\doibase
  10.1038/s41598-019-38777-3} {\bibfield  {journal} {\bibinfo  {journal}
  {Scientific Reports}\ }\textbf {\bibinfo {volume} {9}},\ \bibinfo {pages}
  {2531} (\bibinfo {year} {2019})}\BibitemShut {NoStop}%
\bibitem [{\citenamefont {Kostyukov}\ \emph {et~al.}(2012)\citenamefont
  {Kostyukov}, \citenamefont {Nerush},\ and\ \citenamefont
  {Litvak}}]{lwfa-kostyukov-radiationdamping}%
  \BibitemOpen
  \bibfield  {author} {\bibinfo {author} {\bibfnamefont {I.~Y.}\ \bibnamefont
  {Kostyukov}}, \bibinfo {author} {\bibfnamefont {E.~N.}\ \bibnamefont
  {Nerush}}, \ and\ \bibinfo {author} {\bibfnamefont {A.~G.}\ \bibnamefont
  {Litvak}},\ }\href {\doibase 10.1103/PhysRevSTAB.15.111001} {\bibfield
  {journal} {\bibinfo  {journal} {Phys. Rev. ST Accel. Beams}\ }\textbf
  {\bibinfo {volume} {15}},\ \bibinfo {pages} {111001} (\bibinfo {year}
  {2012})}\BibitemShut {NoStop}%
\bibitem [{\citenamefont {Kimura}\ \emph {et~al.}(2011)\citenamefont {Kimura},
  \citenamefont {Milchberg}, \citenamefont {Muggli}, \citenamefont {Li},\ and\
  \citenamefont {Mori}}]{plasma-kimura-hollow}%
  \BibitemOpen
  \bibfield  {author} {\bibinfo {author} {\bibfnamefont {W.~D.}\ \bibnamefont
  {Kimura}}, \bibinfo {author} {\bibfnamefont {H.~M.}\ \bibnamefont
  {Milchberg}}, \bibinfo {author} {\bibfnamefont {P.}~\bibnamefont {Muggli}},
  \bibinfo {author} {\bibfnamefont {X.}~\bibnamefont {Li}}, \ and\ \bibinfo
  {author} {\bibfnamefont {W.~B.}\ \bibnamefont {Mori}},\ }\href {\doibase
  10.1103/PhysRevSTAB.14.041301} {\bibfield  {journal} {\bibinfo  {journal}
  {Phys. Rev. ST Accel. Beams}\ }\textbf {\bibinfo {volume} {14}},\ \bibinfo
  {pages} {041301} (\bibinfo {year} {2011})}\BibitemShut {NoStop}%
\bibitem [{\citenamefont {Lotov}(1998)}]{pwfa-lotov-selfmodulation}%
  \BibitemOpen
  \bibfield  {author} {\bibinfo {author} {\bibfnamefont {K.~V.}\ \bibnamefont
  {Lotov}},\ }in\ \href
  {https://accelconf.web.cern.ch/AccelConf/e98/PAPERS/MOP12E.PDF} {\emph
  {\bibinfo {booktitle} {Proceedings of 6th European Particle Accelerator
  Conference, Stockholm}}}\ (\bibinfo {year} {1998})\ pp.\ \bibinfo {pages}
  {806--808}\BibitemShut {NoStop}%
\bibitem [{\citenamefont {Kumar}\ \emph {et~al.}(2010)\citenamefont {Kumar},
  \citenamefont {Pukhov},\ and\ \citenamefont
  {Lotov}}]{pwfa-kumar-selfmodulation}%
  \BibitemOpen
  \bibfield  {author} {\bibinfo {author} {\bibfnamefont {N.}~\bibnamefont
  {Kumar}}, \bibinfo {author} {\bibfnamefont {A.}~\bibnamefont {Pukhov}}, \
  and\ \bibinfo {author} {\bibfnamefont {K.}~\bibnamefont {Lotov}},\ }\href
  {\doibase 10.1103/PhysRevLett.104.255003} {\bibfield  {journal} {\bibinfo
  {journal} {Phys. Rev. Lett.}\ }\textbf {\bibinfo {volume} {104}},\ \bibinfo
  {pages} {255003} (\bibinfo {year} {2010})}\BibitemShut {NoStop}%
\bibitem [{\citenamefont {Whittum}\ \emph {et~al.}(1991)\citenamefont
  {Whittum}, \citenamefont {Sharp}, \citenamefont {Yu}, \citenamefont {Lampe},\
  and\ \citenamefont {Joyce}}]{pwfa-whittum-channel}%
  \BibitemOpen
  \bibfield  {author} {\bibinfo {author} {\bibfnamefont {D.~H.}\ \bibnamefont
  {Whittum}}, \bibinfo {author} {\bibfnamefont {W.~M.}\ \bibnamefont {Sharp}},
  \bibinfo {author} {\bibfnamefont {S.~S.}\ \bibnamefont {Yu}}, \bibinfo
  {author} {\bibfnamefont {M.}~\bibnamefont {Lampe}}, \ and\ \bibinfo {author}
  {\bibfnamefont {G.}~\bibnamefont {Joyce}},\ }\href {\doibase
  10.1103/PhysRevLett.67.991} {\bibfield  {journal} {\bibinfo  {journal} {Phys.
  Rev. Lett.}\ }\textbf {\bibinfo {volume} {67}},\ \bibinfo {pages} {991}
  (\bibinfo {year} {1991})}\BibitemShut {NoStop}%
\bibitem [{\citenamefont {Schroeder}\ \emph {et~al.}(1999)\citenamefont
  {Schroeder}, \citenamefont {Whittum},\ and\ \citenamefont
  {Wurtele}}]{pwfa-schroeder-channel}%
  \BibitemOpen
  \bibfield  {author} {\bibinfo {author} {\bibfnamefont {C.~B.}\ \bibnamefont
  {Schroeder}}, \bibinfo {author} {\bibfnamefont {D.~H.}\ \bibnamefont
  {Whittum}}, \ and\ \bibinfo {author} {\bibfnamefont {J.~S.}\ \bibnamefont
  {Wurtele}},\ }\href {\doibase 10.1103/PhysRevLett.82.1177} {\bibfield
  {journal} {\bibinfo  {journal} {Phys. Rev. Lett.}\ }\textbf {\bibinfo
  {volume} {82}},\ \bibinfo {pages} {1177} (\bibinfo {year}
  {1999})}\BibitemShut {NoStop}%
\bibitem [{\citenamefont {Pukhov}\ and\ \citenamefont
  {Farmer}(2018)}]{pwfa-pukhov-coaxialchannels}%
  \BibitemOpen
  \bibfield  {author} {\bibinfo {author} {\bibfnamefont {A.}~\bibnamefont
  {Pukhov}}\ and\ \bibinfo {author} {\bibfnamefont {J.~P.}\ \bibnamefont
  {Farmer}},\ }\href {\doibase 10.1103/PhysRevLett.121.264801} {\bibfield
  {journal} {\bibinfo  {journal} {Phys. Rev. Lett.}\ }\textbf {\bibinfo
  {volume} {121}},\ \bibinfo {pages} {264801} (\bibinfo {year}
  {2018})}\BibitemShut {NoStop}%
\bibitem [{\citenamefont {Pukhov}(2016)}]{pic-pukhov-quasistatic}%
  \BibitemOpen
  \bibfield  {author} {\bibinfo {author} {\bibfnamefont {A.}~\bibnamefont
  {Pukhov}},\ }\href
  {https://e-publishing.cern.ch/index.php/CYR/article/view/220} {\bibfield
  {journal} {\bibinfo  {journal} {CERN Yellow Reports}\ }\textbf {\bibinfo
  {volume} {1}},\ \bibinfo {pages} {181} (\bibinfo {year} {2016})}\BibitemShut
  {NoStop}%
\bibitem [{\citenamefont {Pukhov}(1999)}]{pic-pukhov-vlpl}%
  \BibitemOpen
  \bibfield  {author} {\bibinfo {author} {\bibfnamefont {A.}~\bibnamefont
  {Pukhov}},\ }\href@noop {} {\bibfield  {journal} {\bibinfo  {journal}
  {Journal of Plasma Physics}\ }\textbf {\bibinfo {volume} {61}},\ \bibinfo
  {pages} {425} (\bibinfo {year} {1999})}\BibitemShut {NoStop}%
\bibitem [{\citenamefont {Tarkeshian}\ \emph {et~al.}(2018)\citenamefont
  {Tarkeshian}, \citenamefont {Vay}, \citenamefont {Lehe}, \citenamefont
  {Schroeder}, \citenamefont {Esarey}, \citenamefont {Feurer},\ and\
  \citenamefont {Leemans}}]{plasma-tarkeshian-ionizationprofiling}%
  \BibitemOpen
  \bibfield  {author} {\bibinfo {author} {\bibfnamefont {R.}~\bibnamefont
  {Tarkeshian}}, \bibinfo {author} {\bibfnamefont {J.~L.}\ \bibnamefont {Vay}},
  \bibinfo {author} {\bibfnamefont {R.}~\bibnamefont {Lehe}}, \bibinfo {author}
  {\bibfnamefont {C.~B.}\ \bibnamefont {Schroeder}}, \bibinfo {author}
  {\bibfnamefont {E.~H.}\ \bibnamefont {Esarey}}, \bibinfo {author}
  {\bibfnamefont {T.}~\bibnamefont {Feurer}}, \ and\ \bibinfo {author}
  {\bibfnamefont {W.~P.}\ \bibnamefont {Leemans}},\ }\href {\doibase
  10.1103/PhysRevX.8.021039} {\bibfield  {journal} {\bibinfo  {journal} {Phys.
  Rev. X}\ }\textbf {\bibinfo {volume} {8}},\ \bibinfo {pages} {021039}
  (\bibinfo {year} {2018})}\BibitemShut {NoStop}%
\bibitem [{\citenamefont {Bekov}\ \emph {et~al.}(1977)\citenamefont {Bekov},
  \citenamefont {Letokhov},\ and\ \citenamefont
  {Mishin}}]{plasma-bekov-Rb_ionization}%
  \BibitemOpen
  \bibfield  {author} {\bibinfo {author} {\bibfnamefont {G.}~\bibnamefont
  {Bekov}}, \bibinfo {author} {\bibfnamefont {V.}~\bibnamefont {Letokhov}}, \
  and\ \bibinfo {author} {\bibfnamefont {V.}~\bibnamefont {Mishin}},\ }\href
  {\doibase https://doi.org/10.1016/0030-4018(77)90132-8} {\bibfield  {journal}
  {\bibinfo  {journal} {Optics Communications}\ }\textbf {\bibinfo {volume}
  {23}},\ \bibinfo {pages} {85 } (\bibinfo {year} {1977})}\BibitemShut
  {NoStop}%
\bibitem [{\citenamefont {Balakin}\ \emph {et~al.}(1983)\citenamefont
  {Balakin}, \citenamefont {Novokhatsky},\ and\ \citenamefont
  {Smirnov}}]{pwfa-BNS}%
  \BibitemOpen
  \bibfield  {author} {\bibinfo {author} {\bibfnamefont {V.~E.}\ \bibnamefont
  {Balakin}}, \bibinfo {author} {\bibfnamefont {A.~V.}\ \bibnamefont
  {Novokhatsky}}, \ and\ \bibinfo {author} {\bibfnamefont {V.~P.}\ \bibnamefont
  {Smirnov}},\ }in\ \href
  {http://inspirehep.net/record/198113/files/HEACC83_119-120.pdf} {\emph
  {\bibinfo {booktitle} {Proceedings, 12th International Conference on
  High-Energy Accelerators, HEACC 1983: Fermilab, Batavia, August 11-16,
  1983}}},\ Vol.\ \bibinfo {volume} {C830811}\ (\bibinfo {year} {1983})\ pp.\
  \bibinfo {pages} {119--120}\BibitemShut {NoStop}%
\bibitem [{\citenamefont {Farmer}\ and\ \citenamefont
  {Pukhov}(2019)}]{pwfa-farmer-channel_loading}%
  \BibitemOpen
  \bibfield  {author} {\bibinfo {author} {\bibfnamefont {J.~P.}\ \bibnamefont
  {Farmer}}\ and\ \bibinfo {author} {\bibfnamefont {A.}~\bibnamefont
  {Pukhov}},\ }\href {\doibase 10.1103/PhysRevAccelBeams.22.021003} {\bibfield
  {journal} {\bibinfo  {journal} {Phys. Rev. Accel. Beams}\ }\textbf {\bibinfo
  {volume} {22}},\ \bibinfo {pages} {021003} (\bibinfo {year}
  {2019})}\BibitemShut {NoStop}%
\bibitem [{\citenamefont {Couperus}\ \emph {et~al.}(2017)\citenamefont
  {Couperus}, \citenamefont {Pausch}, \citenamefont
  {K{\~A}{\textparagraph}hler}, \citenamefont {Zarini}, \citenamefont
  {Kr{\~A}{\texteuro}mer}, \citenamefont {Garten}, \citenamefont {Huebl},
  \citenamefont {Gebhardt}, \citenamefont {Helbig}, \citenamefont {Bock},
  \citenamefont {Zeil}, \citenamefont {Debus}, \citenamefont {Bussmann},
  \citenamefont {Schramm},\ and\ \citenamefont {Irman}}]{lwfa-Couperus-loaded}%
  \BibitemOpen
  \bibfield  {author} {\bibinfo {author} {\bibfnamefont {J.~P.}\ \bibnamefont
  {Couperus}}, \bibinfo {author} {\bibfnamefont {R.}~\bibnamefont {Pausch}},
  \bibinfo {author} {\bibfnamefont {A.}~\bibnamefont
  {K\"{o}hler}}, \bibinfo {author} {\bibfnamefont
  {O.}~\bibnamefont {Zarini}}, \bibinfo {author} {\bibfnamefont {J.~M.}\
  \bibnamefont {Kr\"{a}mer}}, \bibinfo {author} {\bibfnamefont
  {M.}~\bibnamefont {Garten}}, \bibinfo {author} {\bibfnamefont
  {A.}~\bibnamefont {Huebl}}, \bibinfo {author} {\bibfnamefont
  {R.}~\bibnamefont {Gebhardt}}, \bibinfo {author} {\bibfnamefont
  {U.}~\bibnamefont {Helbig}}, \bibinfo {author} {\bibfnamefont
  {S.}~\bibnamefont {Bock}}, \bibinfo {author} {\bibfnamefont {K.}~\bibnamefont
  {Zeil}}, \bibinfo {author} {\bibfnamefont {A.}~\bibnamefont {Debus}},
  \bibinfo {author} {\bibfnamefont {M.}~\bibnamefont {Bussmann}}, \bibinfo
  {author} {\bibfnamefont {U.}~\bibnamefont {Schramm}}, \ and\ \bibinfo
  {author} {\bibfnamefont {A.}~\bibnamefont {Irman}},\ }\href {\doibase
  10.1038/s41467-017-00592-7} {\bibfield  {journal} {\bibinfo  {journal}
  {Nature Communications}\ }\textbf {\bibinfo {volume} {8}},\ \bibinfo {pages}
  {487} (\bibinfo {year} {2017})}\BibitemShut {NoStop}%
\end{thebibliography}
%
\end{document}